\documentclass[11pt,a4paper]{article}
\usepackage{jcappub}

\newcommand{\cB}{{\cal B}}
\newcommand{\cE}{{\cal E}}
\newcommand{\cH}{{\cal H}}

\title{Can a marginally open universe amplify magnetic fields?}

\author[a,b]{Yuri Shtanov}
\author[c]{and Varun Sahni}

\affiliation[a]{Bogolyubov Institute for Theoretical Physics,\\ Kiev 03680, Ukraine} %
\affiliation[b]{Department of Physics, Taras Shevchenko National University,\\ Kiev 03022, Ukraine} %
\affiliation[c]{Inter-University Centre for Astronomy and Astrophysics,\\ Post Bag 4,
Ganeshkhind, Pune 411~007, India} %

\emailAdd{shtanov@bitp.kiev.ua}
\emailAdd{varun@iucaa.ernet.in}

\abstract{In a series of recent papers, including arXiv:1210.1183, it was claimed that large-scale
magnetic fields generated during inflation in a spatially open universe could remain
astrophysically significant at the present time since they experienced \emph{superadiabatic
amplification\/} specific to an open universe. We reexamine this assertion
and show that, on the contrary, large-scale magnetic fields in a realistic open universe
decay in much the same manner as they would in a spatially flat universe. Consequently,
their amplitude today is extremely small ($B_0 \lesssim 10^{-59}$~G) and is unlikely to
be of astrophysical significance.}


\arxivnumber{1211.2168}

\begin{document}
\maketitle
\flushbottom

\section{Introduction}

The origin and amplification of galactic and intergalactic magnetic fields constitutes
one of the key open questions in contemporary astrophysics and cosmology
\cite{Grasso:2000wj, Widrow:2002ud, Brandenburg:2004jv, Kandus:2010nw}. Recent
observations \cite{B_voids} of magnetic fields with strengths exceeding $B \sim
10^{-15}$~G in intergalactic space, including underdense regions (voids), has further
enlivened the debate as to whether a dynamo-type mechanism could account for the presence
of such fields, or whether some form of primordial magnetogenesis is required to explain
them.

The present paper will mainly focus on the role of the magnetic field on very large
spatial scales in a marginally open Friedmann universe. In \cite{Barrow:2012ty,
Barrow:2012ax} (see also \cite{Kandus:2010nw, Tsagas:2005nn, Barrow:2008jp, earlier}), it
was argued that the substantially different evolution of the magnetic field in an open
universe (as opposed to a universe which is either spatially flat or closed) may provide
a mechanism by means of which a primordial magnetic field could survive until the present
day, and remain astrophysically relevant. This idea is based on the dynamics of so-called
supercurvature modes of the magnetic field in a spatially open expanding universe, whose
decay rate with the cosmological scale factor $a$ can be slower than the $B \propto
a^{-2}$ decay, typical of a spatially flat universe.

The possibility of exciting supercurvature modes in an open universe remains to be
clarified. Note that these modes do not belong to the space of square-integrable
functions either on hypersurfaces of constant time in an open universe \cite{bd, grib},
or on Cauchy hypersurfaces of the geodesically completed space-time \cite{Sasaki:1994yt,
Adamek:2011hi}. However, they have been used in \cite{Lyth:1995cw} to describe
perturbations with large correlation lengths in an open universe.

In this paper, we show that, even if supercurvature modes of electromagnetic field were
present in a spatially open universe, their evolution would not help in solving the
problem of primordial magnetogenesis in the manner suggested in \cite{Barrow:2012ty,
Barrow:2012ax}. We demonstrate that the behaviour of magnetic fields in a marginally open
universe (at the stage where $1 - \Omega (\eta) \ll 1$) is very similar to its behaviour
in the spatially flat case, and that in both instances $B \propto a^{-2}$ with a high
degree of accuracy. This result is shown to be valid during the exponential stage of open
inflation as well as during the post-inflationary epoch, and holds for the magnetic field
on arbitrarily large spatial scales. In other words, a small value of the spatial
curvature ensures that its effect on the magnetic field is tiny, and implies that a
marginally open universe cannot either preserve or amplify a primordial magnetic field.

\section{The magnetic field in an expanding universe}
\label{sec:magfield}

An expanding homogeneous and isotropic universe is described by the
Friedmann-Robertson-Walker (FRW) metric which, in terms of the conformal time coordinate
$\eta = \int{c dt/a}$, is written as
\begin{equation} \label{metric}
ds^2 = a^2 (\eta) \left[ d \eta^2 - \gamma_{ij} (x) dx^i dx^j \right] \, ,
\end{equation}
where $\gamma_{ij} (x)$ is the spatial homogeneous and isotropic metric, and $c$ is the
speed of light to be set to unity in what follows.

The components $E_i$ and $B_i$ of the observable electric and magnetic fields in such a
universe are conveniently expressed through their conformal counterparts $\cE_i$ and
$\cB_i$ as
\begin{equation} \label{obs}
B_i = \frac{1}{a^2} \cB_i \, , \qquad E_i = \frac{1}{a^2} \cE_i \, .
\end{equation}
A free electromagnetic field obeys the Maxwell equations
\begin{equation} \label{maxwell}
\begin{array}{ll}
{\rm div}\, \vec \cB = 0 \, , \quad & {\rm rot}\, \vec \cE = - {\vec
\cB}' \, , \medskip \\
{\rm div}\, \vec \cE = 0 \, , \quad & {\rm rot}\, \vec \cB = {\vec \cE}'
\, ,
\end{array}
\end{equation}
where a prime denotes partial derivative with respect to the conformal time $\eta$, and
the divergence and rotor operations are defined in the curved three-dimensional space
with the metric $\gamma_{ij}$, associated covariant derivative $\nabla_i$ and normalized
volume element $\epsilon_{ijk}$ as follows:
\begin{equation}
{\rm div}\, \vec a \equiv \nabla_i a^i \, , \qquad \left( {\rm rot}\, \vec a \right)^i
\equiv \epsilon^{ijk} \nabla_j a_k \, .
\end{equation}
All spatial indices are raised and lowered here using the spatial metric $\gamma_{ij}$
and its inverse.

From the Maxwell equations (\ref{maxwell}) one easily obtains a closed equation for the
conformal magnetic field:
\begin{equation}
\cB_{i}'' - \nabla^k \nabla_k \cB_i + R_i{}^k \cB_k = 0 \, ,
\end{equation}
where $R_{ij}$ is the Ricci tensor for the metric $\gamma_{ij}$. In the FRW case, we have
$R_i{}^j = 2 \kappa \delta_i{}^j$, so that the previous equation becomes
\begin{equation}
\cB_i'' - \left(\nabla^k \nabla_k - 2 \kappa  \right) \cB_i = 0 \, .
\end{equation}
Here, $\kappa = 0, \pm 1$ corresponds to the spatial curvature of the homogeneous and
isotropic metric $\gamma_{ij}$.  A similar equation is obtained for the conformal
electric field $\cE_i$.

By separation of time and space variables, the field is decomposed into transverse vector
harmonics which are eigenfunctions of the Laplace operator $\nabla^k \nabla_k$ with
eigenvalues\footnote{In this paper, we follow the notation of \cite{Barrow:2012ty,
Barrow:2012ax} for the eigenvalues of the Laplace operator for transverse vector fields.}
$- n^2$, and, for the coefficients $\cB_{(n)} (\eta)$ of this decomposition, one obtains
the ordinary differential equation
\begin{equation} \label{harmon}
\cB_{(n)}'' + \left( n^2 + 2 \kappa \right) \cB_{(n)} = 0 \, .
\end{equation}

Restricted to the space of square-integrable vector functions, the Laplace operator has a
continuum spectrum with $n^2 > - 2 \kappa$ in the flat or open geometry, and a discrete
spectrum with $n^2 = p^2 - 2$, $p = 2,3, \ldots$, in the closed geometry (see, e.g.,
\cite{Goode:1989jt}). As can be seen from (\ref{harmon}), all modes in this spectrum have
harmonic oscillatory solutions
\begin{equation}\label{modes}
\cB_{(n)} = C_1 \cos{\left( \eta \sqrt{n^2 + 2 \kappa}\right)}
 + C_2 \sin{\left( \eta \sqrt{n^2 + 2 \kappa}\right)}
\, ,
\end{equation}
where $C_1$ and $C_2$ are integration constants, so that the amplitudes of components
(\ref{obs}) of the observable magnetic (and electric) field decay as $B \propto a^{-2}$.

In the case of a spatially open universe,  $\kappa = - 1$, the solutions of
(\ref{harmon}) with $n^2 < 2$ are usually called \emph{supercurvature modes\/} (our
definition of supercurvature modes matches that in \cite{Adamek:2011hi}, while only modes
with $0 \leq n < 1$ are referred to as being supercurvature in \cite{Barrow:2012ty}).  As
functions of the radial coordinate, such modes do not oscillate but exhibit purely
hyperbolic behaviour;\footnote{The corresponding scalar radial mode characterized by the
number $n$ and by the angular number $l$ has the behaviour \cite{Sasaki:1994yt,
Adamek:2011hi}
$$
f_{nl}(r) \propto \sinh^l r \frac{d^{\,l}}{d \left(\cosh r \right)^l} \left( \frac{\sin
\sqrt{n^2 - 2}\, r}{\sinh r} \right) \, .
$$
For $n^2 > 2$, the modes oscillate with comoving wavelength $\lambda = 2 \pi / \sqrt{n^2
- 2}$.  For $n^2 < 2$, they exhibit purely hyperbolic behaviour. Transverse vector modes
are constructed from the scalar modes by additional differentiation \cite{Adamek:2011hi}.
\label{remark}} in this sense, they cannot be properly characterized by a wavelength.
Although the supercurvature modes do not belong to the space of square integrable
functions, their scalar counterparts were used to describe perturbations with large
correlation lengths in a spatially open cosmological model \cite{Lyth:1995cw}.

As follows from (\ref{modes}), the supercurvature vector modes also have a qualitatively
different temporal behaviour\footnote{In \cite{Barrow:2008jp}, the difference between
(\ref{modes}) and (\ref{cbn}) was attributed to the fact that an open FRW universe is
locally (but not globally) conformal to the flat Minkowski space \cite{Iihoshi:2007uz}.}
\begin{equation} \label{cbn}
\cB_{(n)} = C_1 e^{\eta\,\sqrt{2 - n^2}} + C_2 e^{- \eta\,\sqrt{2 - n^2}} \, .
\end{equation}
The growing mode in (\ref{cbn}) allows us to relate the final value of the magnetic
field, $\cB_{(n)}^{({\rm f})}$, to its initial value, $\cB_{(n)}^{({\rm i})}$, as
follows:
\begin{equation}\label{cbn1}
\frac{\cB_{(n)}^{({\rm f})}}{\cB_{(n)}^{({\rm i})}}  = e^{\alpha ( \eta_{\rm f} - \eta_{\rm i} )} \equiv
e^{\alpha\Delta\eta} \, ,
\end{equation}
where $\Delta\eta = \eta_{\rm f} - \eta_{\rm i}$ and $\alpha = \alpha (n) = \sqrt{2 -
n^2}$. This equation implies that the temporal change in $\cB_{(n)}$ depends upon the
time span $\Delta \eta$. At this point, we should emphasize that, since $a (\eta) d \eta
= dt$ determines the physical cosmological time, the scale of the conformal time is
uniquely fixed by the normalization of the scale factor $a$.  In a spatially closed or
open universe, it is fixed by the conventional choice $\kappa = \pm 1$ for the spatial
curvature of the metric $\gamma_{ij}$ in (\ref{metric}). With this choice, both the scale
factor $a$ and the scale of the conformal time $\eta$ in (\ref{metric}) have absolute
geometrical, and therefore also physical, meaning.

Two extreme cases will be of interest to us in this paper: (a)~$\Delta \eta \gg 1$, since
this limit played a key role in the deductions of \cite{Barrow:2012ty, Barrow:2012ax};
(b)~$\Delta \eta \ll 1$, which we show to provide a more accurate description of the
phase of exponential inflation as well as the post-inflationary epoch. As one can see from
(\ref{cbn1}), during epochs spanning a large range of values of the conformal time,
namely $\alpha\Delta \eta \gg 1$, the asymptotic growth of supercurvature modes of
$\cB_{(n)}$ is exponential:
\begin{equation} \label{expon}
\cB_{(n)} \simeq \cB_{(n)}^{({\rm i})}e^{\alpha (\eta - \eta_{\rm i})}
\end{equation}
whereas in the opposite case, when $\Delta \eta \ll 1$, the field $\cB_{(n)}$ freezes at
a constant value, $\cB_{(n)}^{({\rm f})} \simeq \cB_{(n)}^{({\rm i})}$, implying $ B
\propto a^{-2}$ even for supercurvature modes.

\section{Large values of the conformal time}

A homogeneous and isotropic universe with metric (\ref{metric}) is described by the
Friedmann equation
\begin{equation} \label{friedmann}
\cH^2 + \kappa = \frac{8 \pi G}{3} a^2 \sum_i\rho_i \, ,
\end{equation}
together with the conservation equation for the density component $\rho_i$:
\begin{equation} \label{state}
\rho_i' + 3 \cH (\rho_i + p_i) = 0 \, .
\end{equation}
Here, $\cH \equiv a' / a = a H$ is the conformal version of the Hubble parameter $H
\equiv \dot a / a$, and an overdot denotes the derivative with respect to the physical
time $t$. Of special significance will be the curvature parameter for such a universe:
\begin{equation} \label{omk}
\Omega_\kappa (\eta) = 1 - \Omega (\eta) \equiv - \frac{\kappa}{a^2 H^2} = -
\frac{\kappa}{\cH^2} \, .
\end{equation}

An exact solution of (\ref{friedmann}), (\ref{state}) for a spatially open universe
($\kappa = -1$) filled with matter with constant parameter of equation of state $w \equiv
p/\rho$ is easily found to be (see also \cite{Barrow:2012ty, earlier})
\begin{equation} \label{a}
a (\eta) = a_* \left( \frac{\sinh \beta \eta}{\sinh \beta \eta_*} \right)^{1/\beta} \, ,
\end{equation}
where
\begin{equation} \label{beta}
\beta = \frac{1 + 3 w}{2} \neq 0 \, ,
\end{equation}
and $\eta_*$ and $a_* = a (\eta_*)$ are integration constants. For the expansion law
(\ref{a}), one finds
\begin{equation}
\cH = \coth \beta \eta \, ,
\end{equation}
and the curvature parameter (\ref{omk}) has the form
\begin{equation} \label{eq:om_curv}
\Omega_\kappa (\eta) = \tanh^2 \beta \eta \, ,
\end{equation}
so that a large absolute value of $\eta$ in (\ref{eq:om_curv}) implies a large value of
$\Omega_\kappa (\eta)$, and, conversely, a small value of $\Omega_\kappa (\eta)$ implies
a small absolute value for $\eta$.

Two toy models played a key role in the magnetic-field analysis of \cite{Barrow:2012ty,
Barrow:2008jp}, namely:
\begin{itemize}
\item[(i)] $a \propto (\sinh{\eta})^{-1}$, which describes open inflation and
    corresponds to $w = -1 \ \Rightarrow \ \beta = -1$ in
    (\ref{a})--(\ref{eq:om_curv}). \label{one}
\item[(ii)] $a \propto \sinh{\eta}$, which describes an open radiation-dominated
    universe and corresponds to $w = 1/3 \ \Rightarrow \ \beta = 1$ in
    (\ref{a})--(\ref{eq:om_curv}). \label{two}
\end{itemize}

The value of the conformal time $\eta$ in (\ref{a}) is calculated from the past
cosmological singularity ($a = 0$) if $\beta > 0$, and from the future asymptotic
infinity ($1/a = 0$) if $\beta < 0$.  In both cases, for a large span of the conformal
time parameter, namely, for $| \beta| \Delta \eta \gg 1$, the scale factor (\ref{a})
evolves exponentially with $\eta$,
\begin{equation}\label{asymptote}
a (\eta) \propto e^{\eta} \, ,
\end{equation}
and describes the empty Milne universe. For (i) the Milne asymptote precedes exponential
inflation (with $a (\eta) \propto - 1/\eta \propto e^{Ht}$) and therefore lies in the
remote past, whereas for (ii) the Milne asymptote succeeds a radiation-dominated epoch
and therefore lies in the remote future.\footnote{Solution (\ref{a}) played an important
role in the conclusions drawn in \cite{Barrow:2012ty, Barrow:2012ax}. However, this
framework is somewhat simplistic since the universe has several components, including
dark energy. As we shall show in the next section, the asymptotic expansion law
(\ref{asymptote}) characteristic of an empty Milne universe is an oversimplification
which is never achieved in the real universe with the cosmological constant.} The
presence of the Milne asymptote is quite general for arbitrary $\beta$, or even for a
spatially open universe filled with arbitrary matter, and arises whenever the total
density of this matter is subdominant to the spatial curvature in (\ref{friedmann}),
i.e., when
\begin{equation}
\frac{8 \pi G}{3} a^2 \sum_i \rho_i \ll 1 \, ,
\end{equation}
and $\Omega_\kappa (\eta) \simeq 1$ in (\ref{omk}).

During the curvature-dominated regime described by (\ref{asymptote}), the evolution
(\ref{cbn}) of the supercurvature modes of the magnetic field can be presented as
\begin{equation} \label{cba}
\cB_{(n)} = C_1 a^{\sqrt{2 - n^2}} + C_2 a^{- \sqrt{2 - n^2}}\, ,
\end{equation}
so that
\begin{equation} \label{ba}
B_{(n)} = C_1 a^{\sqrt{2 - n^2} - 2} + C_2 a^{- \sqrt{2 - n^2} - 2}\, ,
\end{equation}
and the first mode of the observable magnetic field in (\ref{ba}) decays considerably
less rapidly than given by the usual law $B \propto a^{-2}$.  This effect was highlighted
in a number of papers \cite{Kandus:2010nw, Barrow:2012ty, Barrow:2012ax, Tsagas:2005nn,
Barrow:2008jp, earlier} and led to the claim that it remained valid during ``a period of
slow-roll inflation, during reheating, and subsequently in the radiation and dust
epochs,'' and that throughout this time large-scale $B$-fields were
``\emph{superadiabatically amplified\/}\footnote{The \emph{superadiabatic
amplification\/} of the electromagnetic field by curvature effects should be
distinguished from the superadiabatic amplification of quantum fields which takes place
during inflation. In the case of the latter, the amplitude of a given mode that leaves
the Hubble radius during inflation is superadiabatically amplified relative to a much
higher momentum mode which never left the Hubble radius and whose amplitude therefore
decreased adiabatically throughout. Superadiabatic amplification in quantum language
translates into particle production, and prominent examples of this process include the
inflationary production of gravity waves \cite{grishstar} and fields that couple
non-minimally to gravity \cite{Widrow:2002ud, Turner:1987bw, Opher:1997jn, Sahni:1998at}.
Note that this does not happen in the case of the electromagnetic field, which couples
conformally to gravity and whose modes, therefore, remain in the vacuum state throughout
the expansion of a FRW space-time \cite{Widrow:2002ud, bd, grib}. } by curvature effects
alone'' \cite{Barrow:2012ax}. As we have seen, the Milne asymptote (\ref{asymptote}), so
very vital for the derivation of (\ref{cba}) and (\ref{ba}), and the basis for the claim
that supercurvature modes of the $B$-field are superadiabatically amplified, demands
$\Omega_\kappa (\eta) \simeq 1$. As we proceed to show in the next section, the
transition from eq.~(\ref{cbn}) to eqs.~(\ref{cba}) and (\ref{ba}), which is precisely
the transition from eq.~(5) to eq.~(6) in ref.~\cite{Barrow:2012ax}, is valid only at a
possible early stage of cosmological expansion dominated by spatial curvature (`coasting
phase' in the terminology of \cite{Barrow:2012ty}) but is erroneous at the subsequent
stage of exponential inflation and also after inflation, during which $\Omega_\kappa
(\eta) \ll 1$ and the magnetic field decays as $B \propto a^{-2}$.

\section{Small values of the conformal time}
\label{sec:small}

The authors of \cite{Barrow:2012ty, Barrow:2012ax, Tsagas:2005nn, Barrow:2008jp, earlier}
used expressions (\ref{cba}) and (\ref{ba}) during the inflationary as well as
post-inflationary (radiation and matter dominated) stages to sustain magnetic fields on
large spatial scales in supercurvature modes.  However, as we have already noted, the
asymptotic expressions (\ref{cba}) and (\ref{ba}) were obtained under the assumption of
the exponential behaviour (\ref{asymptote}) of the scale factor, which is valid only when
the conformal time parameter in (\ref{a}) spans a large range of values. Unfortunately,
this condition can be valid only in a curvature-dominated universe, and does not apply to
the phase of exponential inflation and post-inflationary epoch, during which both
$\Omega_\kappa (\eta)$ and $\Delta \eta$ are small. Let us consider this issue in more
detail.

\subsection{Magnetic-field evolution during open inflation}
\label{sec:inflat}

Cosmological inflation is usually driven by an ingredient (a scalar field) with effective
equation of state $w \approx - 1$.  Setting $w = - 1$ in (\ref{a}) and (\ref{beta})
results in the following expansion law for open inflation:
\begin{equation} \label{ainf}
a (\eta) = a_* \frac{\sinh \eta_*}{\sinh \eta} \, .
\end{equation}
In this conventional expression, the conformal time $\eta$ is counted from the future
asymptotic infinity (where $1/a = 0$).  It spans negative values and is decreasing by
absolute magnitude as the universe expands. Since $\cH = - \coth{\eta}$, the deceleration
parameter in such a universe,
\begin{equation} \label{q}
q \equiv - \frac{\ddot a}{aH^2} = -\frac{\cH'}{\cH^2}  =
-\frac{1}{\cosh^2 \eta} \, ,
\end{equation}
describes the `coasting' (Milne) phase $a (t) \propto t$ when $| \eta| \gg 1$ with $|q|
\ll 1$, and the phase of exponential inflation $a (t) \propto e^{Ht}$ when $| \eta | \ll
1$ (with $|q| \to 1$ as $\eta \to 0$). The fact that the absolute value of $\eta$ is
small during inflation can also be seen from the expression for the curvature parameter
(\ref{eq:om_curv}). Setting $\beta = -1$ in (\ref{eq:om_curv}), we obtain
\begin{equation} \label{omtinf}
\Omega_\kappa (\eta) = \tanh^2 \eta \, ,
\end{equation}
which reflects the geometrical meaning of the value of the conformal time, as noted in
the previous section. It is sensible to agree that inflation `commences' when the
inflaton energy density begins to dominate the spatial curvature. This occurs when the
value of the curvature parameter has dropped below $\Omega_\kappa = 0.5$. From
(\ref{omtinf}) we find the corresponding value of $\eta$ to be $|\eta| \approx 0.88$.
Hence, $|\eta| < 1$ during inflation.

Well within the regime of exponential inflation, one has $| \eta | \ll 1$, and
(\ref{ainf}) reduces to the usual flat-space behaviour
\begin{equation}
a( \eta) \approx a_* \frac{\sinh \eta_*}{\eta} \, .
\end{equation}
Equation (\ref{cbn}) in this case takes the form
\begin{equation}\label{Bflat}
\cB_{(n)} \approx \tilde C_1 + \tilde C_2\, \eta\, \sqrt{2 - n^2} \simeq \tilde C_1
\, ,
\end{equation}
where $\tilde C_1 = C_1 + C_2$ and $\tilde C_2 = C_1 - C_2$. In other words, since the
span of conformal time during exponential inflation is small ($\Delta \eta \lesssim 1$),
the field $\cB_{(n)}$ tends to a constant value and the physical magnetic field
asymptotically decays as $B \propto a^{-2}$. This leads us to conclude that no
`superadiabatic amplification' of the magnetic field occurs during exponential inflation.

The previous reasoning was based on the de~Sitter solution (\ref{ainf}). However, a
generic estimate can be made for the span $\Delta \eta$ of the conformal time in any
model of open quasi-exponential inflation, satisfying the condition $| \dot H | \ll H^2$,
between the moments of its beginning and its end:
\begin{equation}
\Delta \eta = \int_{t_{\rm in}}^{t_{\rm end}} \frac{d t}{a (t)} = \int_{a_{\rm
in}}^{a_{\rm end}} \frac{d a}{a^2 H} \simeq \frac{1}{a_{\rm in} H_{\rm in}} =
\sqrt{\Omega_\kappa^{\rm in}} \, .
\end{equation}
Since, in any case, $\Omega_\kappa^{\rm in} < 1$, we get an upper bound $\Delta \eta
\lesssim 1$.

It is instructive to estimate the value of $\Omega_\kappa$ at the end of inflation (or at
the commencement of reheating, which, for simplicity, is assumed to start immediately
after the end of inflation):
\begin{equation} \label{ominf}
\Omega_\kappa (\eta_{\rm end}) \simeq \Omega_\kappa (\eta_{\rm rh}) \simeq \left(
\frac{g_0}{g_{\rm rh}} \right)^{1/3} \left( \frac{T_0}{T_{\rm rh}} \right)^2
\frac{\Omega_{0\kappa}}{\Omega_{0\gamma}} \simeq 10^{-52} \left( \frac{10^3}{g_{\rm rh}}
\right)^{1/3} \left( \frac{10^{15}\, \mbox{GeV}}{T_{\rm rh}} \right)^2 \Omega_{0\kappa}
\, .
\end{equation}
Here, $T_{\rm rh}$ is the temperature of reheating, $T_0 \approx 2.7~\mbox{K} \simeq 2.3
\times 10^{-4}~\mbox{eV}$ is the current temperature of the cosmic microwave background
(CMB), while $\Omega_{0\gamma} \approx 5 \times 10^{-5}$ is the current value of the CMB
energy density. The quantity $g_0 = 2$ is the number of degrees of freedom of the photon
and $g_{\rm rh}$ is the number of relativistic degrees of freedom in thermal equilibrium
after reheating. The current value of the curvature parameter in the $\Lambda$CDM model
is constrained by observations \cite{Komatsu:2010fb} to lie in the interval $- 0.0133 <
\Omega_{0\kappa} < 0.0084$ (95\%~CL), so it is conceivable that ${\Omega_{0\kappa}}
\simeq 10^{-2}$, which was suggested in \cite{Barrow:2012ty, Barrow:2012ax} to describe a
marginally open universe. Substituting ${\Omega_{0\kappa}} = 10^{-2}$ in (\ref{ominf}),
we obtain ${\Omega_\kappa (\eta_{\rm end})} \approx \eta_{\rm end}^2 \simeq 10^{-54}$ at
the end of inflation. Consequently, $| \eta | \simeq 10^{-27}$ when inflation ended, and
$| \eta | \sim 1$ when exponential inflation commenced (about 62 $e$-foldings before the
end of inflation). Thus, $| \eta |$ remains small during the entire duration of
exponential inflation, and we arrive at the conclusion that (\ref{Bflat}), rather than
(\ref{cba}), provides the correct description of the behaviour of the magnetic field
during inflation in an open universe.

\subsection{The magnetic field in the post-inflationary universe}
\label{sec:post}

As previously noted, a key role in the derivation of  superadiabatic amplification of
large-scale $B$-modes in (\ref{cba}) and (\ref{ba}) was played by the assumption that the
cosmological scale factor displays behaviour characteristic of the curvature-dominated
Milne universe, namely $a \propto e^\eta$. However, as we have seen in a previous
subsection, the behaviour of the scale factor during exponential inflation is $a (\eta)
\propto - 1 / \eta$ with $\eta \to 0^-$, so that no superadiabatic amplification occurs
at this stage.

In \cite{Barrow:2008jp}, it has been claimed that the superadiabatic amplification of the
$B$ field is not confined to the inflationary phase but extends also to the radiation and
matter dominated epochs which follow preheating. For this to be the case, the expansion
law $a \propto e^\eta$ needs to be valid during radiation and matter domination. In this
section, we show that this is not the case and that the Milne asymptote $a \propto
e^\eta$, characteristic of a curvature-dominated empty universe, is never followed during
post-inflationary expansion.  The reason of this behaviour is quite similar to that
discussed in the previous subsection in connection with exponential inflation: during
post-inflationary expansion, the physically relevant conformal time $\eta$ spans a small
range of values $\Delta \eta \ll 1$, commencing from exceedingly small initial values and
remaining small until today. As a consequence, the scale factor $a (\eta)$ behaves as a
power of $\eta$, rather then as an exponent.

A spatially open universe filled with radiation and matter is described by  the following
exact solution of the Friedmann equation (\ref{friedmann}) \cite{Mukhanov:1990me}:
\begin{equation} \label{amrad}
a (\eta) = a_{\rm eq} \left( \zeta \sinh \eta + \zeta^2 \sinh^2 \frac{\eta}{2}
\right) \, , \qquad \eta > 0 \, ,
\end{equation}
where
\begin{equation}
\zeta = \left( \frac{8 \pi G}{3} \rho_{\rm eq} a_{\rm eq}^2 \right)^{1/2} = \left( \frac{
1 - \Omega_{\kappa\, \rm eq}}{2 \Omega_{\kappa\, \rm eq}} \right)^{1/2} = \left[
\frac{\Omega_{\rm eq}}{2 \left( 1 - \Omega_{\rm eq} \right)} \right]^{1/2} \, ,
\end{equation}
the subscript `eq' refers to the moment of equality of matter and radiation densities,
and $\rho_{\rm eq}$ is the density of matter or radiation at that moment.  The conformal
time at this stage of expansion is conventionally counted from the extrapolated
cosmological singularity ($a = 0$ as $\eta = 0$).

The curvature parameter (\ref{omk}) for solution (\ref{amrad}) is given by\footnote{Only
the first term in the brackets of (\ref{amrad}) was retained in \cite{Barrow:2008jp} for
a description of a radiation-dominated universe, which is somewhat misleading because
matter plays a singularly more important role than spatial curvature during this epoch. Indeed,
taking the cosmological matter parameter $\Omega_{0m} \simeq 1/3$ and $\Omega_{0\kappa}
\simeq 10^{-2}$, we have $\frac{\Omega_\kappa(z)}{\Omega_m(z)} =
\frac{\Omega_{0\kappa}}{\Omega_{0m}} (1 + z)^{-1}$, so that the ratio of curvature to
matter becomes a minuscule $10^{-5}$ at $z \sim 10^4$, and was even smaller earlier on.}
\begin{equation} \label{omcur}
\Omega_\kappa (\eta) = \left( \frac{\sinh \eta + \zeta \sinh^2 \frac{\eta}{2}}{\cosh
\eta + \zeta \cosh \frac{\eta}{2} \sinh \frac{\eta}{2}} \right)^2 \, ,
\end{equation}
from which we find that the exceedingly small value of $\Omega_\kappa (\eta)$ during the
radiation and matter dominated epochs implies $\eta \ll 1$.  (In fact, by assuming, for
simplicity, instantaneous reheating and matching the parameter $\Omega_\kappa$ at the end
of inflation and at reheating, we immediately get the relation $\eta_{\rm rh} \simeq |
\eta_{\rm end}|$ between the values of the conformal times at reheating and at the end of
inflation.) Such a small value of $\eta$ ensures that equation (\ref{amrad}) reduces to
its flat-space counterpart
\begin{equation}\label{a_composite}
a (\eta) \approx a_{\rm eq} \left( \zeta \eta + \frac14 \zeta^2 \eta^2 \right) \, ,
\end{equation}
and the evolution (\ref{cbn}) of the conformal magnetic field is approximated as
\begin{equation}\label{B_rad}
\cB_{(n)} \approx \tilde C_1 + \tilde C_2 \sqrt{2 - n^2}\, \eta \simeq \tilde C_1 \,
,
\end{equation}
implying $B \propto a^{-2}$. We, therefore, conclude that the large-$\eta$ asymptote $a
\propto e^\eta$ used in the derivation of superadiabatic amplification in (\ref{cba}) is
inapplicable, and the more relevant equations (\ref{a_composite}) and (\ref{B_rad}) imply
that the magnetic field decreases in much the same manner as it does in a spatially flat
universe, namely $B \propto a^{-2}$.

Equation (\ref{amrad}), which takes into account the effect of spatial curvature but
misses the effect of dark energy, accurately describes the real universe only until dark
energy starts dominating over the spatial curvature. In the $\Lambda$CDM model, this
takes place at redshift $z = \sqrt{\Omega_\Lambda / \Omega_{0\kappa}} - 1 \approx 7.5$
for $\Omega_\Lambda = 0.73$ and $\Omega_{0\kappa} = 0.01$. After the cosmological
constant starts dominating absolutely, we proceed to an asymptotically de~Sitter stage
again. Therefore, during the post-inflationary expansion of a $\Lambda$CDM universe, the
curvature term is always strongly subdominant. Consequently, the Milne expansion law
$a\propto e^\eta$, which was crucial in the transition from eq.~(\ref{cbn}) to
eqs.~(\ref{cba}) and (\ref{ba}) made in \cite{Barrow:2012ty, Barrow:2012ax,
Tsagas:2005nn, Barrow:2008jp, earlier}, is never (even asymptotically) valid in a
$\Lambda$CDM universe.

Since the numerical value of the conformal time has played a key role in the above
discussion, we give an exact formula for the span of $\eta$ in a spatially open
post-inflationary Friedmann universe.  The Hubble parameter has the form
\begin{equation}
h(z) \equiv \frac{H (z)}{H_0} = \left[ \Omega_{0r} (1 + z)^4 + \Omega_{0m} (1 + z)^3 +
\Omega_{0\kappa} (1 + z)^2 + \Omega_\Lambda  \right]^{1/2} \, ,
\end{equation}
where we assume for simplicity that the role of dark energy is played by the cosmological
constant and take into account that the contribution to pressureless matter comes from
baryons as well as dark matter. The value of the curvature parameter at an earlier epoch
is related to its present value as
\begin{equation} \label{eq:omk}
\Omega_k(z) = \Omega_{0\kappa} \frac{(1+z)^2}{h^2(z)} \, ,
\end{equation}
while the conformal time coordinate as a function of redshift has the form
\begin{equation} \label{eq:eta1}
\eta (z) = \frac{1}{a_0H_0}\int_z^\infty\frac{d x}{h (x)} \, ,
\end{equation}
which reduces to
\begin{equation} \label{eq:eta2}
\eta (z) = \sqrt{\Omega_{0\kappa}}\int_z^\infty\frac{d x}{h(x)}
\end{equation}
in a spatially open universe.

A simple calculation shows that the value of $\eta$ was small in the past and shall remain
small also in the future. Note that, for $z \geq 0$, the integral in
(\ref{eq:eta2}) is bounded by its present value:
\begin{equation}
\int_z^\infty \frac{d z}{h (z)} \leq \int_0^\infty \frac{d z}{h (z)} \simeq 3.4 \, .
\end{equation}
Substituting $\Omega_{0\kappa} = 10^{-2}$ in (\ref{eq:eta2}), therefore, implies $\eta(z)
< 0.34$ for $z > 0$. The total span of the conformal time in the $\Lambda$CDM model is
obtained by integrating from $z = -1$ in (\ref{eq:eta2}), with the result $\Delta \eta =
\eta (-1) \approx 0.45$ for $\Omega_{0\kappa} = 10^{-2}$.  These results are borne out by
an exact determination of $\eta(z)$ whose values are shown in table~\ref{tab:table1} for
different cosmological redshifts, assuming present-day cosmological parameters which are
consistent with the data analysis of the Wilkinson Microwave Anisotropy Probe
\cite{Komatsu:2010fb}.

\begin{table}[ht]
\begin{center}
\begin{tabular}{|c|c|c|}
\hline \hline
$z$    &  $\Omega_\kappa (z)$           &  $\eta(z)$     \\
\hline
   $-1$         &  0  &  0.45 \\
   0        &  0.01  &  0.34 \\
   10      &  $3.5 \times 10^{-3}$  &  0.11 \\
   3200     &  $7.4 \times 10^{-6}$  &  $2.7 \times 10^{-3}$     \\
$10^9$ & $2 \times 10^{-16}$ & $1.4 \times 10^{-8}$ \\
\hline \hline
\end{tabular}
\end{center}
\caption{\label{tab:table1} The spatial-curvature parameter $\Omega_\kappa (z)$ in
(\ref{eq:omk}) and the conformal time coordinate $\eta(z)$ in (\ref{eq:eta2}) are shown
for typical values of the cosmological redshift.}
\end{table}

\section{The magnetic field on large spatial scales}

In the previous section, we have  demonstrated the absence of superadiabatic
amplification of the magnetic field at the stage of exponential inflation and after
inflation. In view of these results, we would like to revise the estimates of
\cite{Barrow:2012ty, Barrow:2012ax} for possible current values of the magnetic field.
Considering the modes with $1 < n^2 < 2$, the authors of \cite{Barrow:2012ty,
Barrow:2012ax} specified their initial amplitudes at the respective moments of time
defined by the condition $n^2 = \cH^2 (\eta_{\rm HC})$ and termed `horizon crossing.'
Since the vector modes under consideration do not exhibit oscillatory behaviour either in
the radial coordinate or in time, and, therefore, cannot be characterized by a real
wavelength or frequency (see the discussion in section \ref{sec:magfield} and footnote
\ref{remark}), this characterization of `horizon crossing' appears somewhat artificial.
Nevertheless, following \cite{Barrow:2012ty, Barrow:2012ax}, we assume these initial
values for the magnetic-field amplitudes and trace their evolution in an open universe.
Even in this case, as we are going to show, our results turn out to be quite different
from those derived in \cite{Barrow:2012ty, Barrow:2012ax}.

During the epoch of open inflation, described by (\ref{ainf}), the condition of `horizon
crossing' $n^2 = \cH^2 (\eta_{\rm HC})$ translates into
\begin{equation} \label{cross}
n^2 = \coth^2 \eta_{\rm HC} \quad \Rightarrow \quad \eta_{\rm HC} = - \coth^{-1} n
\end{equation}
(remember that $\eta < 0$ at this stage). Note that the relation (\ref{omk}) implies
\begin{equation} \label{omhc}
\Omega_k(\eta_{\rm HC}) = \frac{1}{n^2} \, ,
\end{equation}
which informs us that the moment of `horizon crossing' for modes with lower $n$ occurs
when the value of $\Omega_k$ is larger (see also \cite{Barrow:2012ty}).

The magnetic-field modes under consideration evolve according to (\ref{cbn}). Since our
aim is to see what maximal possible magnetic field one can obtain today, we retain only
the growing mode in this equation. At the initial moment of `horizon crossing,' we then
have
\begin{equation}
\cB_{\rm HC} = C_1 e^{\eta_{\rm HC} \sqrt{2 - n^2}} = C_1 \left(\frac{n - 1}{n + 1}
\right)^{\frac{\sqrt{2 - n^2}}{2}} \, .
\end{equation}
Towards the end of inflation, we have $|\eta_{\rm end}| \lesssim 10^{-27} \ll 1$ (see
section \ref{sec:inflat}), which leads to
\begin{equation} \label{ampl1}
\cB_{\rm end} \simeq C_1 \approx \left(\frac{n + 1}{n - 1} \right)^{\frac{\sqrt{2 -
n^2}}{2}} \cB_{\rm HC} \, .
\end{equation}
Since $B = \cB/a^2$, one obtains the following relation for physical magnetic fields:
\begin{equation} \label{ampl2}
B_{\rm end} \simeq \left( \frac{a_{\rm HC}}{a_{\rm end}} \right)^2 \left(\frac{n +
1}{n - 1} \right)^{\frac{\sqrt{2 - n^2}}{2}} B_{\rm HC} \, .
\end{equation}

As we have shown in section \ref{sec:post}, during post-inflationary era, the magnetic
field in all modes decays approximately as $B \propto a^{-2}$.  Consequently, for the
present value of the magnetic field, one gets
\begin{equation} \label{Btoday}
B_0 = \left( \frac{a_{\rm end}}{a_0} \right)^2 B_{\rm end} \approx \left( \frac{a_{\rm
HC}}{a_0} \right)^2 \left(\frac{n + 1}{n - 1} \right)^{\frac{\sqrt{2 - n^2}}{2}}
B_{\rm HC} \, .
\end{equation}
The ratio of the scale factors in (\ref{Btoday}) is determined using (\ref{omtinf}) and
(\ref{cross}), and, by keeping in mind the extreme smallness of $| \eta_{\rm end}|$, we
have
\begin{equation}
\frac{a_{\rm HC}}{a_{\rm end}} = \frac{\sinh \eta_{\rm end}}{\sinh \eta_{\rm HC}} =
- \sqrt{n^2 - 1}\, \sinh \eta_{\rm end} \approx \sqrt{(n^2 - 1) \Omega_\kappa^{\rm end}}
\, ,
\end{equation}
\begin{equation}
\frac{a_{\rm end}}{a_0} = \frac{a_{\rm end} H_{\rm end}}{a_0 H_0} \cdot \frac{H_0}{H_{\rm
end}} = \sqrt{\frac{\Omega_{0\kappa}}{\Omega_\kappa^{\rm end}}} \cdot \frac{H_0}{H_{\rm
end}} \, ,
\end{equation}
so that
\begin{equation}
\frac{a_{\rm HC}}{a_0} = \frac{a_{\rm HC}}{a_{\rm end}} \cdot \frac{a_{\rm
end}}{a_0} = \sqrt{(n^2 - 1) \Omega_{0\kappa}}\, \frac{H_0}{H_{\rm end}} \, .
\end{equation}
Substituting this into (\ref{Btoday}), we get
\begin{equation} \label{main}
B_0 \simeq (n + 1)^{\sqrt{2 - n^2}} \left( n^2 - 1 \right)^{\frac{2 - \sqrt{2 -
n^2}}{2}}\, \Omega_{0\kappa} \frac{H_0^2}{H_{\rm end}^2} B_{\rm HC} \, .
\end{equation}

If, following \cite{Barrow:2012ty, Tsagas:2005nn}, we assume that $B_{\rm HC} \simeq
H_{\rm end}^2$, then, from (\ref{main}), we obtain\footnote{Note that the assumption
$B_{\rm HC} \simeq H_{\rm end}^2$ made in \cite{Barrow:2012ty, Tsagas:2005nn} is somewhat
speculative since hard calculations are missing in support of this point of view.
Moreover, Adamek et al.\@ \cite{Adamek:2011hi} have recently drawn attention to an
inherent difficulty in exciting supercurvature modes in open inflation and in the Milne
universe.}
\begin{eqnarray} \label{final}
B_0 &\simeq& (n + 1)^{\sqrt{2 - n^2}} \left( n^2 - 1 \right)^{\frac{2 -
\sqrt{2 - n^2}}{2}}\, \Omega_{0\kappa} H_0^2 \nonumber \\
&\simeq& 10^{-63} (n + 1)^{\sqrt{2 - n^2}} \left( n^2 - 1 \right)^{\frac{2 - \sqrt{2 -
n^2}}{2}}\, \Omega_{0\kappa} h^2\, {\rm G} \, ,
\end{eqnarray}
where $h = H_0 \left/ \left( 100~\mbox{km s$^{-1}$ Mpc$^{-1}$} \right) \right.$.
Remarkably, the final value of the magnetic field in (\ref{final}) does not depend upon
the energy scale of inflation. Equation (\ref{final}) should be compared with formula
(35) of \cite{Barrow:2012ty}:
\begin{equation} \label{barrow}
B_0 \sim 10^{- 65 + 51 \sqrt{2 - n^2}} \left( \frac{M}{10^{14}\, \mbox{GeV}} \right)^{2
\sqrt{2 - n^2}} \left[ (n^2 - 1) \Omega_{0\kappa} \right]^{\frac{2 - \sqrt{2 -
n^2}}{2}}\, {\rm G} \, ,
\end{equation}
where $M$ is the energy scale of inflation.  The estimate for $B_0$ in (\ref{barrow})
differs by many orders of magnitude from the correct estimate in (\ref{final}) mainly
because the authors of \cite{Barrow:2012ty} employed the asymptotic formulas (\ref{cba})
and (\ref{ba}) not only at the coasting phase, where they are valid, but also at the
inflationary stage and during post-inflationary evolution, where they are, in fact,
inapplicable.

Our result (\ref{main}) gives a very small estimate (\ref{final}) for the magnetic field
on supercurvature scales today. Even by increasing the energy density stored in the magnetic
field during `horizon crossing' to the inflationary energy density, we get $B_{\rm HC}
\sim M_{\rm P} H_{\rm end} \sim 10^5 H_{\rm end}^2$ (here, $M_{\rm P}$ is the Planck
mass), which, while increasing our estimate (\ref{final}) by five orders of magnitude,
still leaves it very small.

\section{A bound on the free magnetic field in a marginally open universe}

It is widely believed that, since Maxwell's equations couple conformally to gravity, a
primordial magnetic field cannot be created solely by the expansion of a FRW universe,
but might do so if conformal invariance were somehow broken. Indeed, several attempts at
magnetogenesis have introduced explicit couplings of electromagnetism either to gravity
\cite{Turner:1987bw}, or to the inflaton \cite{Ratra:1991bn}, to break conformal
invariance and generate primordial magnetic fields.

As this paper was not directly concerned with the problem of magnetogenesis, we did not
feel it appropriate to discuss this issue. Rather, since we were mainly concerned with
revisiting some of the issues raised in \cite{Barrow:2012ty, Tsagas:2005nn}, we simply
assumed, as did the authors of \cite{Barrow:2012ty, Tsagas:2005nn}, that the seed value
of the magnetic field was linked to the Hubble value during inflation. One may, however,
broaden the scope of the above discussion by asking whether an upper bound can be placed
on the value of a non-interacting (free) primordial magnetic field by requiring that it
remain compatible with open inflation.

Let us assume, for instance, that an open universe was created as a quantum bubble with
some prevailing (seed) magnetic field.  Initially, the energy density in such a universe
may be dominated either by spatial curvature, or by the magnetic field, or by some other
form of matter.  We assume, however, that the universe at some moment of time $\eta_{\rm
in}$ begins to inflate. After this moment, its Hubble expansion is dominated by the
energy density of the inflaton, hence, both the magnetic-field energy density and the
spatial curvature are subdominant.  Such a universe quickly approaches the regime where
it is adequately described by the FRW solution (\ref{ainf}).

As we have shown in section \ref{sec:small}, all modes of the free magnetic field decay
approximately as $B \propto a^{-2}$ both during and after inflation.  Since the magnetic
field density was subdominant to that of the inflaton at the start of inflation, and
since the free magnetic field does not experience any \emph{additional\/} amplification
(also true for the supercurvature modes, as discussed in section \ref{sec:small}),  we
obtain a simple relation between the energy density contained in these modes at the
beginning and at the end of exponential inflation:
\begin{equation}  \label{mag}
\rho^B_{\rm end} \simeq \left( \frac{a_{\rm in}}{a_{\rm end}} \right)^4 \rho^B_{\rm in} =
e^{- 4 N} \rho^B_{\rm in} < e^{- 4 N} \rho_{\rm in} \, ,
\end{equation}
where $\rho_{\rm in}$ is the initial energy density of the inflaton and $N$ is the number
of $e$-foldings during exponential inflation. For the current value of the magnetic
field, we get
\begin{eqnarray}
B_0 &\sim& \left( \frac{a_{\rm end}}{a_0} \right)^2 B_{\rm end} \sim \left( \frac{a_{\rm
end}}{a_0} \right)^2  e^{- 2 N} \sqrt{\rho^B_{\rm in}} < \left( \frac{a_{\rm end}}{a_0}
\right)^2  e^{- 2 N} \sqrt{\rho_{\rm in}} \nonumber \\ &\sim& g_{\rm rh}^{-1/6} \left(
\frac{\rho_{\rm in}}{\rho_{\rm end}} \right)^{1/2} e^{- 2 N} T_0^2 \simeq 2.5 \times
10^{-7} \left( \frac{10^3}{g_{\rm rh}} \right)^{1/6} \left( \frac{\rho_{\rm
in}}{\rho_{\rm end}} \right)^{1/2} e^{- 2 N}~{\rm G} \, .
\end{eqnarray}
The lower bound $N > 60$ on the number of $e$-foldings in the simplest inflationary
models leads to the following upper bound on the present value of the magnetic field on
supercurvature scales: $B_0 \lesssim 10^{-59}$~G\@. A significantly larger magnetic
field, say, $B_0 \sim 10^{- 16}$~G, would imply $N < 11$ and run into trouble with our
current understanding of inflation.

\section{Discussion}

In this paper, we have investigated the evolution of supercurvature modes of the magnetic
field in a marginally open universe.  We have shown that, contrary to the claims made in
\cite{Barrow:2012ty, Barrow:2012ax}, such modes do not experience any significant
amplification either during exponential inflation or during the post-inflationary epoch.
The basic reason for this is that, while the conformal field $\cB = a^2 B$ in these modes
does evolve exponentially with conformal time, the conformal time itself during these stages
spans a very small range of values, $\Delta \eta \ll 1$, (while the scale factor $a
(\eta)$ evolves as a power of $\eta$ and not as $a \propto e^\eta$).
Thus, the conformal magnetic field $\cB$ remains
effectively frozen in time, and the physical magnetic field $B$ decays as $a^{-2}$.
Following the evolution of the magnetic field in open inflation with
initial conditions as in \cite{Barrow:2012ty, Tsagas:2005nn}, we arrive at a very
small estimate (\ref{final}) for its current value:
$B_0 \lesssim 10^{-65}$~G.

By considering the contribution of the magnetic field to the rate of expansion of the
universe, one can obtain a general upper bound on the residual free magnetic field in an
open universe, compatible with a sufficiently long epoch of exponential inflation. For $N
= 60$ inflationary $e$-foldings, we arrive at the estimate $B_0 \lesssim 10^{- 59}$~G\@.

The possibility of exciting supercurvature modes of the magnetic field in an open
inflationary universe was put into question in \cite{Adamek:2011hi} basically because
these modes do not belong to the space of square-integrable functions
\cite{Sasaki:1994yt, Adamek:2011hi}. In this paper, the important issue of excitation of
supercurvature modes of the electromagnetic field has been set aside.  Instead, we have
shown that even if these modes were somehow present, their current amplitude in a
marginally open universe would be too small to account for primordial magnetogenesis.

\acknowledgments

We are grateful to Alexei Starobinsky and Kandaswamy Subramanian for valuable comments.
The authors acknowledge support from the India-Ukraine Bilateral Scientific Cooperation
programme. Yu.~S\@. was also supported by the Cosmomicrophysics section of the Programme
of the Space Research of the National Academy of Sciences of Ukraine and by the Swiss
National Science Foundation (SCOPES grant no.~128040).

\end{document}